\documentclass[12pt]{article}
\usepackage{amsmath}
\usepackage{graphicx}
\usepackage{times}
\usepackage[T1]{fontenc}
\usepackage[utf8]{inputenc}
\usepackage[margin=1cm]{caption}

\usepackage[square,numbers,sort]{natbib}

\author
{Tauqir Shinwari$^{1,\dag,\ast}$, Kacho Imtiyaz Ali Khan$^{1,\dag,\ast}$, Hua Lv$^{1}$,  Atekelte Abebe Kassa$^{1}$,
\\Frans Munnik$^{2}$, Simon Josephy$^{3}$, Achim Trampert$^{1}$, Victor Ukleev$^{4}$, Chen Luo$^{4}$, \\Florin Radu$^{4}$, Jens Herfort$^{1}$, 
Michael Hanke$^{1}$ and Joao Marcelo Jordao Lopes$^{1,\ast}$
\\
\normalsize{$^{1}$Paul-Drude-Institut für Festkörperelektronik, Leibniz-Institut im Forschungsverbund Berlin e.V.,}\\
\normalsize{Berlin, Germany.}\\
\normalsize{$^{2}$Helmholtz-Zentrum Dresden-Rossendorf, Institute of Ion Beam Physics and Materials Research e.V.,}\\
\normalsize{Dresden, Germany.}\\
\normalsize{$^{3}$QZabre AG, Neunbrunnenstrasse 50,}
\normalsize{8050 Zürich, Switzerland}\\
\normalsize{$^{4}$Helmholtz Zentrum Berlin for Materialien und Energie, Albert-Einstein Straße 15,}\\
\normalsize{12489 Berlin, Germany}\\
\\
\normalsize{$^\dag$These authors contributed equally to this work.} \\
\normalsize{$^\ast$To whom correspondence should be addressed; E-mails:} \\
\normalsize{shinwari@pdi-berlin.de, khan@pdi-berlin.de, lopes@pdi-berlin.de}
}

\date{}

\begin{document}

\parindent 0cm
\parskip 12pt

\title{\LARGE\bfseries{Above-room-temperature ferromagnetism in large-area epitaxial Fe$_3$GaTe$_2$/graphene van der Waals heterostructures}} 

\maketitle

\begin{abstract}
Fe$_{3}$GaTe$_{2}$ (FGaT), a two-dimensional (2D) layered ferromagnetic metal, exhibits a high Curie temperature ($T_{\rm C}$) $\sim$ 360 K along with strong perpendicular magnetic anisotropy (PMA), making it a promising material candidate for next-generation energy-efficient magnetic devices. However, the vast majority of studies on FGaT to date have been limited to millimeter-sized bulk crystals and exfoliated flakes, which are unsuitable for practical applications and integration into device processing. Also, its combination with other 2D materials to form van der Waals heterostructures has only been achieved by flake stacking. Consequently, the controlled large-scale growth of FGaT and related heterostructures remains largely unexplored. In this work, we demonstrate a breakthrough in the high-quality, large-scale growth of epitaxial FGaT thin films on single-crystalline graphene/SiC templates using molecular beam epitaxy. Structural characterization confirms the high crystalline quality of the continuous FGaT/graphene van der Waals heterostructures. Temperature-dependent magnetization and anomalous Hall measurements reveal robust PMA with an enhanced $T_{\rm C}$ well above room temperature, reaching up to 400 K. Furthermore, X-ray absorption and X-ray magnetic circular dichroism spectra provide insight into the spin and orbital magnetic moment contributions, further validating the high $T_{\rm C}$ and robust PMA. These findings are highly significant for the future development of high-performance spintronic devices based on 2D heterostructures, with potential applications in next-generation data storage, logic processing and quantum technologies.
\end{abstract}

\section{Introduction}
The emergence of two-dimensional (2D) van der Waals (vdW) magnetic materials has opened new frontiers in fundamental physics, in particular in the exploration of exotic magnetic phenomena such as topologically non-trivial spin textures in the 2D limit.~\cite{gibertini_2019, wang_magnetic_2022} They have also been shown to be promising building blocks for the realization of novel devices for spintronics and quantum technologies with low-power consumption.~\cite{yang_two-dimensional_2022, cabo2025roadmap} Here, a particularly promising approach is to combine such layered magnets with other 2D crystals (e.g., graphene, WSe$_2$) to create vdW heterostructures with multiple functionalities and properties that can be tailored via heterostructure design and proximity-induced phenomena.~\cite{sierra_van_2021, lin_recent_2023} 
Moreover, 2D magnetic heterostructures are advantageous over conventional metallic magnetic heterostructures,~\cite{zabel2007magnetic,shinwari2021bulk,torrejon2014interface,gelen2024growth,ikeda2010perpendicular,shinwari2023growth} as their ultrathin nature is expected to facilitate the realization of ultra-compact devices for efficient charge and spin transport.~\cite{zhang20242d,lemme20222d,novoselov20162d,xue2024recent,chakraborty2022challenges}
In the last couple of years, the library of 2D vdW magnets has expanded rapidly.~\cite{kurebayashi_magnetism_2022} Several (anti)ferromagnetic materials possessing a semiconductor or metallic nature have been identified.~\cite{zhang20242d} However, only a few of them (typically ferromagnetic metals) exhibit transition temperatures around or above room temperature,~\cite{lv_large-area_2023, zhang_above-room-temperature_2022,  chen_pervasive_2022} which is essential for their implementation in a wide range of applications. Among these materials, the 2D ferromagnetic metal Fe$_{3}$GaTe$_{2}$ (FGaT) has recently gained significant attention due to its exceptional magnetic properties. In addition to having one of the highest reported Curie temperatures (between 350 K and 380 K),~\cite{zhang_above-room-temperature_2022, lee_electronic_2023} FGaT also possesses a strong perpendicular magnetic anisotropy (PMA), which is crucial for applications in magnetic memories,~\cite{yang_two-dimensional_2022,kajale_current-induced_2024, kajale_field-free_2024} sensors,~\cite{huang_giant_2025, li_spinorbit_2021} and skyrmion-based logic devices.~\cite{li_room-temperature_2024, ahn_2d_2020} 

\begin{figure}[t!]
		\begin{center}
			\includegraphics[width=0.92\columnwidth]{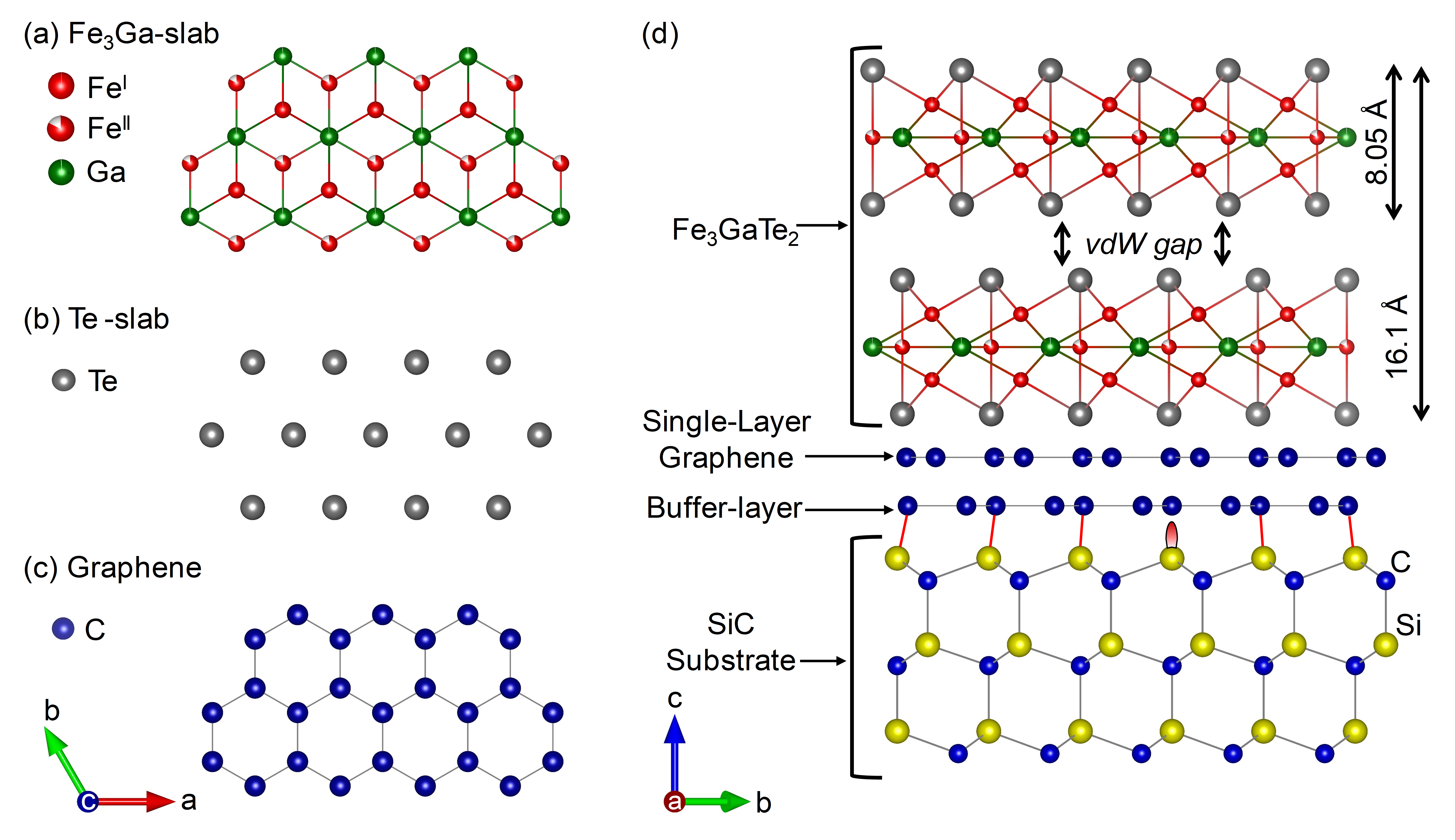}
			\caption { Schematics of the structural configuration of the synthesized FGaT/graphene/SiC(00.1) system.~\cite{fromm_contribution_2013} (a) Top view of the inner Fe$_{3}$Ga slab, where the hexagonal arrangement can be clearly observed. (b) Configuration of the Te atoms forming the outer slabs which encapsulate Fe$_{3}$Ga. (c) Atomic structure of a graphene monolayer. (d) Side view along the [2-1.0] crystallographic orientation of the FGaT layer, which is grown on top of the epitaxial graphene/SiC(00.1) template. Note that the latter also contains a carbon buffer layer at the interface between graphene and SiC, which is covalently bonded to SiC. The axis labels shown in the bottom left and right are related to (a,b,c) and (d), respectively. }.
			\label{fig0}	
		\end{center}
\end{figure}

FGaT, which is a close analogue to the vdW ferromagnet Fe$_{3}$GeTe$_{2}$ (FGT), shows a hexagonal symmetry with a space group $P6_3/mmc$ (No. 194) and bulk lattice parameters $a$ = $b$ = 4.09~\AA, $c$ = 16.07~\AA.~\cite{lee_electronic_2023} The atomic structure of a FGaT single layer consists of a Fe$_{3}$Ga slab which is sandwiched between two Te slabs [Fig.~{\ref{fig0}}(a), (b)]. Adjacent layers are held together by weak vdW forces [Fig.~{\ref{fig0}}(d)], which allows the exfoliation of micrometer-sized thin flakes from bulk single crystals.~\cite{wang_hard_2024,zhang_above-room-temperature_2022, algaidi_magnetic_2024, kajale_current-induced_2024} Both bulk crystals and flakes have been extremely important for studying the intrinsic properties of FGaT, either alone~\cite{zhang_above-room-temperature_2022, wu_spectral_2024, lv_distinct_2024} or in combination with other 2D materials via assembly of vdW heterostructures.~\cite{huang_giant_2025, xiao_polarity-reversal_2024, jin_room-temperature_2023, pan_roomtemperature_2024} However, they are inherently incompatible with device fabrication processes. Hence, to implement FGaT and related heterostructures into future applications, it is necessary to develop large-scale growth of FGaT as crystalline thin films with properties comparable to, or even superior to, those reported for state-of-the-art bulk crystals and flakes. Furthermore, the ability to synthesize FGaT directly on another functional 2D material, ideally without the use of layer transfer, is also highly demanded to enable large-scale vdW heterostructures with ultra-clean interfaces, which is critical for the future realization of novel devices where interface-related proximity effects will play a critical role.~\cite{zheng_lateral_2024, pandey_energy-efficient_2024}

In this work, we demonstrate the large-scale growth of FGaT epitaxial thin films directly on single-crystalline graphene [on SiC(00.1)], see [Fig.~{\ref{fig0}}(c) and (d)], via molecular beam epitaxy (MBE). Continuous and uniform FGaT/graphene vdW heterostructures exhibiting sharp vdW interfaces could be realized, which is a significant advance with respect to what can be achieved via conventional flake stacking. Importantly, magnetization and magneto-transport characterization performed with different methods reveal the FGaT films on graphene to display robust PMA and high $T_{\rm C}$ values approaching 400 K. These results represent a significant step towards the development of ultra-compact, all 2D spintronic devices, such as spin valves~\cite{zhao_room-temperature_nodate, zheng_lateral_2024} and spin de-multiplexers,~\cite{sierra_van_2021} which can be fabricated using vdW heterostructures formed of FGaT and graphene and operate at room temperature. 

\section{Results and Discussion}
\subsection{Structural properties}
\begin{figure}[b!]
\centering
\includegraphics[width=0.92\linewidth]{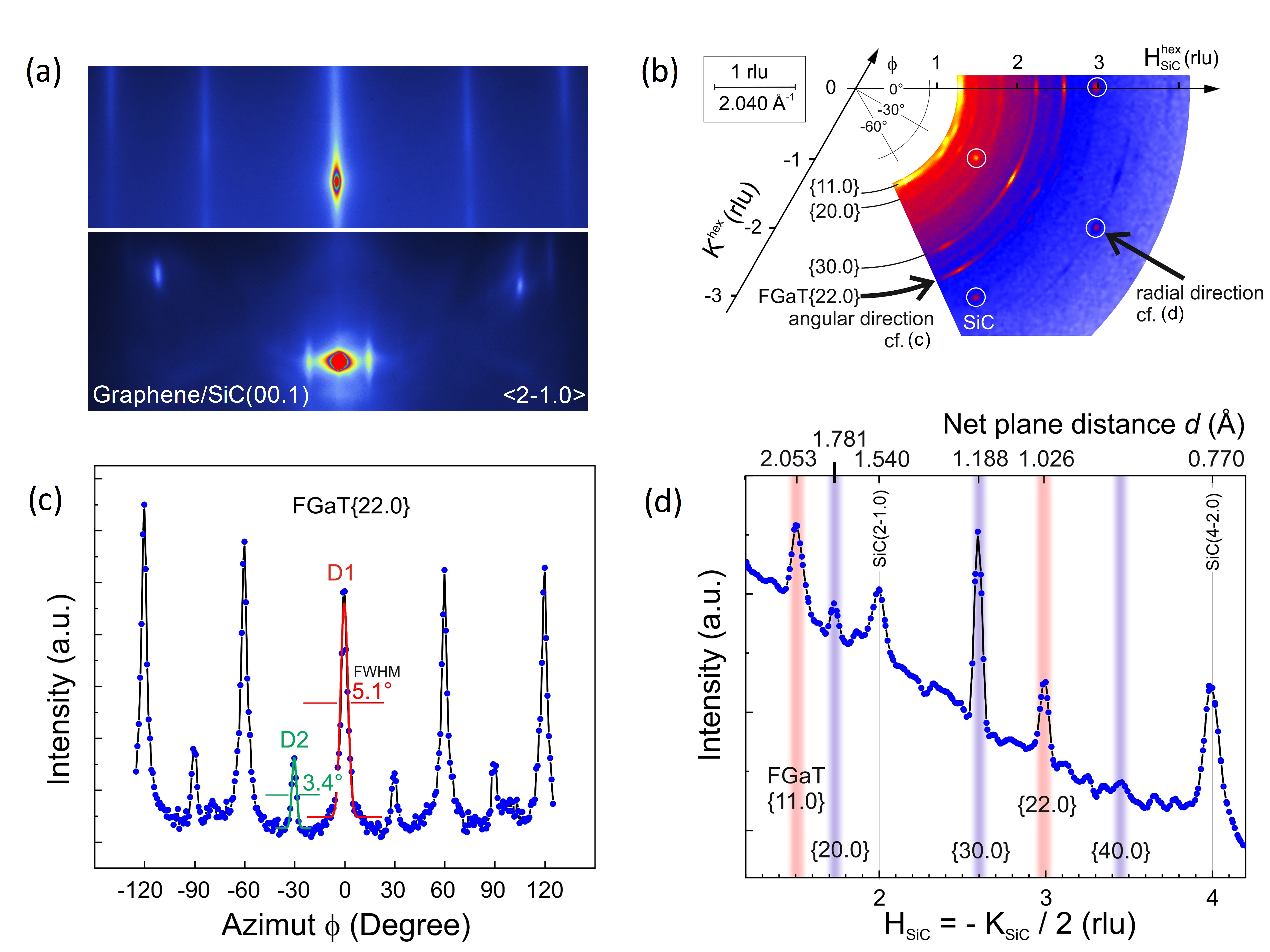}
\caption{\label{fig1}(a) In-situ RHEED patterns obtained for the graphene/SiC template [bottom panel] and after the growth of a 10 nm thick FGaT film [upper panel]. Both patterns were taken perpendicular to the <2-1.0> direction of the SiC(00.1) substrate. (b) GID in-plane reciprocal space map of a 10 nm thick FGaT film on graphene/SiC(00.1). In addition to well-defined reflections from the SiC substrate (white circles), contributions from the FGaT film can be identified. (c) Azimuthal scan of the FGaT[22.0] peak along the angular direction. (d) The radial direction peaks corresponding to FGaT[11.0], [22.0], as well as FGaT[20.0], [30.0], [40.0] are denoted by red and purple vertical lines, respectively, as indicated in the in-plane reciprocal map (b).}
\end{figure}
MBE was utilized to realize a high-quality growth of FGaT films with thicknesses of 6, 10, and 32 nm on graphene/SiC(00.1) templates, which were prepared by SiC surface graphitization.~\cite{heilmann_influence_2020} Rutherford backscattering spectrometry (see Fig.~S1 in the supplementary file) revealed the films to exhibit an average chemical composition close to Fe$_{3.15}$GaTe$_{1.47}$, namely Fe$_{\rm3+ x}$GaTe$_{\rm2-y}$. The FGaT growth process was continuously monitored by in-situ RHEED. The bottom and upper panels in Fig.~\ref{fig1}(a)  correspond to the RHEED patterns of the graphene/SiC(00.1) template and of a $\sim$10 nm thick FGaT film, respectively. The RHEED pattern of the FGaT film shows sharp and continuous streaks, confirming the successful epitaxial growth of a FGaT 2D film having a smooth surface. Based on the position of the streaks, we could extract the in-plane lattice parameters of the FGaT film to be  $a\sim$4.08~\AA. The in-plane structural properties were investigated in more detail by utilizing synchrotron-based grazing incidence diffraction (GID) measurements. This measurement allows us to extend the areas in reciprocal space and provide a comprehensive view of the existing and diffracting in-plane lattice planes and their azimuthal distribution, as shown in Fig.~\ref{fig1}(b)-(d). One particular feature of this technique refers to the fact that the diffraction vector contains exclusively in-plane components (i.e., the vertical component practically equals zero because of the grazing incidence and exit). In other words, GID approaches the ideal case of a non-coplanar geometry where the surface normal $\textbf{n}$ appears perpendicular to the scattering plane. Consequently, a 2D GID reciprocal space map, as in Fig.~\ref{fig1}(b), shows the scattered X-ray intensity parallel to the sample surface. It can be seen as a compiled sequence of concentric arcs (around the reciprocal origin and probing along $q_{\rm angular}$) with increasing diameter, given by $q_{\rm radial}$.  In addition to well-defined reflections from the SiC substrate (white circles), there are other contributions due to the FGaT layer. Most of them show a particular in-plane arrangement with respect to the substrate; e.g., following the arc due to FGaT$\{22.0\}$ net planes, one can see pronounced maxima every 60° and weaker ones in between [Fig.~\ref{fig1}(c)]. This shows a preferential arrangement in which FGaT[11.0] is parallel to SiC[10.0] and, complementary FGaT[10.0] $\parallel$ SiC[11.0]. The areas below the red (D1) and green (D2) curve fittings in  Fig.~\ref{fig1}(c) give a ratio of about 4:1 for the bimodal domain distribution. Fig.~\ref{fig1}(d) shows a line plot in the radial direction intersecting two SiC substrate peaks, namely SiC(2-1.0) and the following (4-2.0) of the same family. For both orientations of the film there are at least two reflection orders, i.e., $\{11.0\}$ and $\{22.0\}$ and $\{n0.0\}$, \textit{n}=2...4, which allow for a precise determination of the in-plane lattice parameter of the MBE-grown FGaT film to be $a = 4.108[7]$~\AA. This value is in excellent agreement with our RHEED value and also consistent with reported values for bulk crystals~\cite{lee_electronic_2023, chen_lattice_2025} and MBE-grown films on mica.~\cite{wu_robust_2024}

\begin{figure}[t!]
	\begin{center}
			\includegraphics[width=0.92\columnwidth]{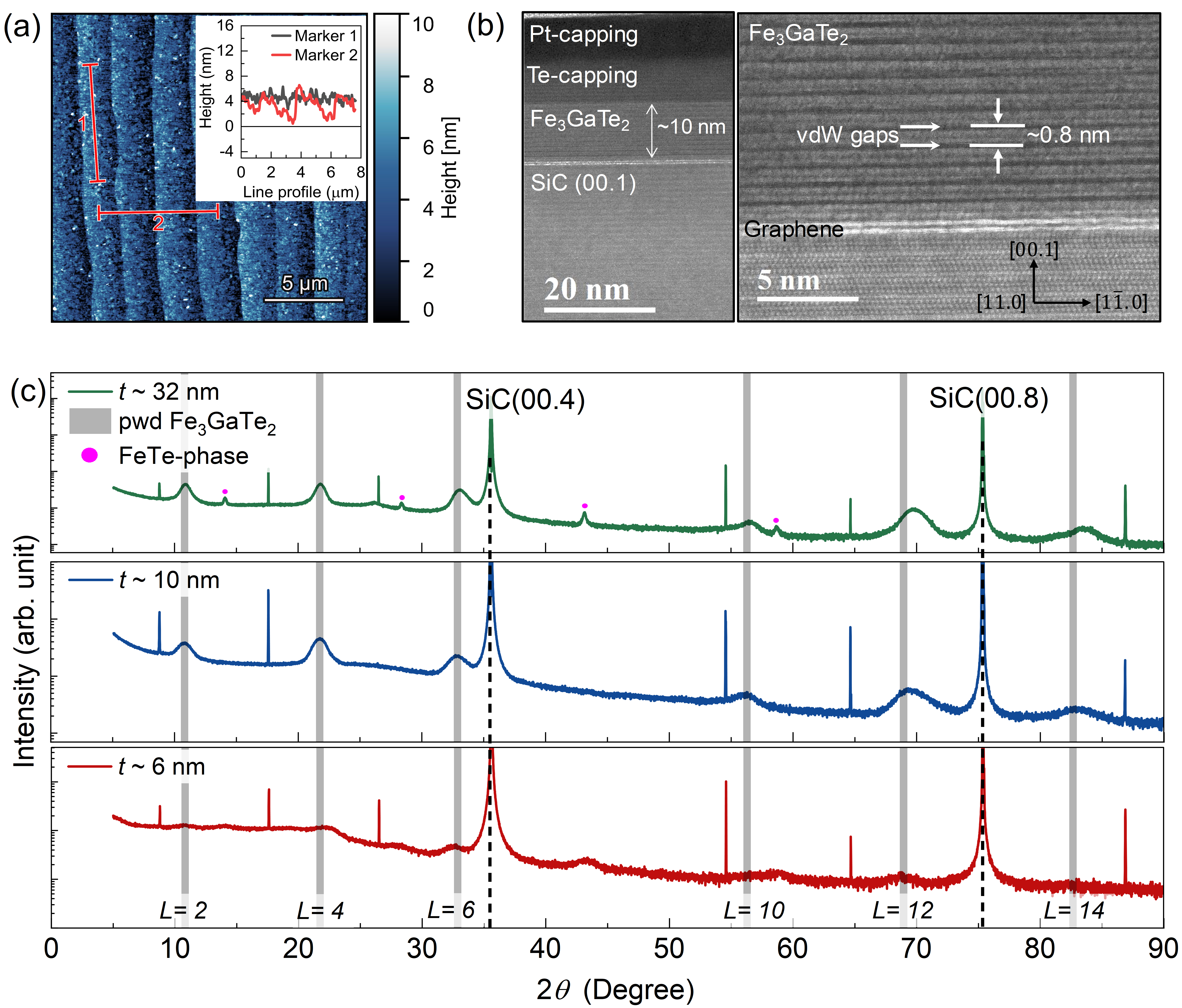}
			\caption {(a) AFM height image of a 10~nm thick FGaT grown on graphene/SiC(00.1). The FGaT film is capped with a ~ $\sim$7 nm thick Te. The inset shows the line profiles corresponding to the markers 1 and 2. (b) Bright-field STEM images for $\sim$10~nm thick FGaT film obtained along the [$11.0$] direction. While the left-side image shows the whole stacking, including the Pt/Te cappings, the right-side image shows a magnified region where it is possible to see the layered structure of FGaT and graphene. (c) The $\theta-2\theta$ scans for for the 6~nm, 10~nm, and 32~nm thick FGaT films containing a series of reflections, whose positions are in good agreement with the calculated powder (pwd) pattern for FGaT (gray bars).  The red dots indicate small contribution from the tetragonal FeTe phase, which forms for the 32 nm thick FGaT case. The vertical dash line represents the Bragg reflection for SiC substrate.}
			\label{fig2}	
		\end{center}
\end{figure}

Fig.~\ref{fig2}(a) shows an atomic force microscopy (AFM) image of a 10~nm thick FGaT film capped with a 7~nm thick Te layer. The line profiles displayed in the inset corresponding to the markers 1 and 2 in Fig.~\ref{fig2}(a) represent the topography of the film along and perpendicular to the terrace step edges of graphene/SiC(00.1). A small value of root-mean-square (rms) roughness of 0.69 nm (along the line profile for marker 1) for this film indicates the growth of the FGaT film with very good uniformity without significant thickness variations or island formation.  Also, no discernible discontinuities or pinholes in the film, e.g., close to step edges in graphene/SiC(00.1), could be observed. This is in contrast with our earlier reports on FGT grown on graphene/SiC(00.1),~\cite{lopes_large-area_2021} where such imperfections were more prevalent. Such observed morphological characteristics suggest a highly controlled growth process, resulting in  FGaT films that homogeneously cover the underlying graphene/SiC template, including regions close to step edges where transition between monolayer to bi-(few-)layer graphene is normally present.~\cite{lopes_large-area_2021} This uniform film structure is crucial for maintaining consistent electronic and magnetic properties across the sample, which in turn is essential for further device processing and thus applications.

The structure of the FGaT/graphene heterostructures was also assessed via cross-sectional scanning transmission electron microscopy (STEM). Fig.~\ref{fig2}(b) shows bright-field images of a 10 nm thick FGaT film on graphene/SiC(00.1), which was capped with Pt/Te layers. The left-side image shows the smooth interface of the FGaT film to both the Te capping layer and the graphene/SiC(00.1) substrate, while the higher magnification image on the right side clearly reveals the uniform layered nature of the FGaT films (in that single layers are separated by vdW gaps) and its sharp interface to the underlying graphene. The thickness of each FGaT quintuple layer [formed by sequential Te/Fe/FeGa/Fe/Te slabs, see Fig.~\ref{fig0}(a)] is determined to be $\sim$ 0.8 nm, which is in agreement with values obtained for FGaT bulk crystals.~\cite{lee_electronic_2023} 

XRD $\theta-2\theta$ scans were performed to probe the out-of-plane structure of the FGaT films [Fig.~\ref{fig2}(c)] grown on the graphene/SiC(00.1) substrate. The presence of (00.\textit{L}) reflections (where \textit{L} = 2, 4, ..., 14) for 6 nm and 10 nm thick FGaT indicates the formation of a pure crystalline FGaT phase. However, the presence of additional small reflections (marked by a magenta dots) for the case of 32~nm thick film revealed the formation of tetragonal FeTe in addition to the FGaT phase, with its c-axis aligned parallel to the SiC [00.1] direction. Tetragonal FeTe, which is antiferromagnetic with a N\'{e}el temperature of about 70 K, has also been observed in MBE-grown Fe$_{5-x}$GeTe$_{2}$ films.~\cite{lv_large-area_2023} The reason for its formation is not well understood and needs further investigation. The Bragg peak positions and relative intensities of the FGaT reflections closely matched the numerically calculated powder diffraction pattern (indicated by grey bars), corroborating the structural integrity of the FGaT films. The out-of-plane lattice parameters were determined using the Debye-Scherrer formula~\cite{holzwarth_scherrer_2011} and it is found to be around 16.39~\r{A}, 16.41~\r{A}, and 16.30~\r{A} for the 6~nm, 10~nm, and 32~nm thick FGaT films, respectively, which is in excellent agreement with our STEM results [note that a single FGaT layer corresponds to 1/2 of the unit cell; see Fig.~\ref{fig0}(a)]. These values are also consistent with previously reported results for both FGaT bulk crystals~\cite{zhang_above-room-temperature_2022} and FGaT thin films.~\cite{wu_robust_2024} It is noteworthy that these XRD measurements were performed on Te-capped FGaT films, which may have implications for the observed lattice parameters and peak intensities. The presence of the Te capping layer could potentially influence the strain state of the underlying FGaT film, particularly in the case of ultrathin samples.

\begin{figure}[t!]
		\begin{center}
			\includegraphics[width=0.92\columnwidth]{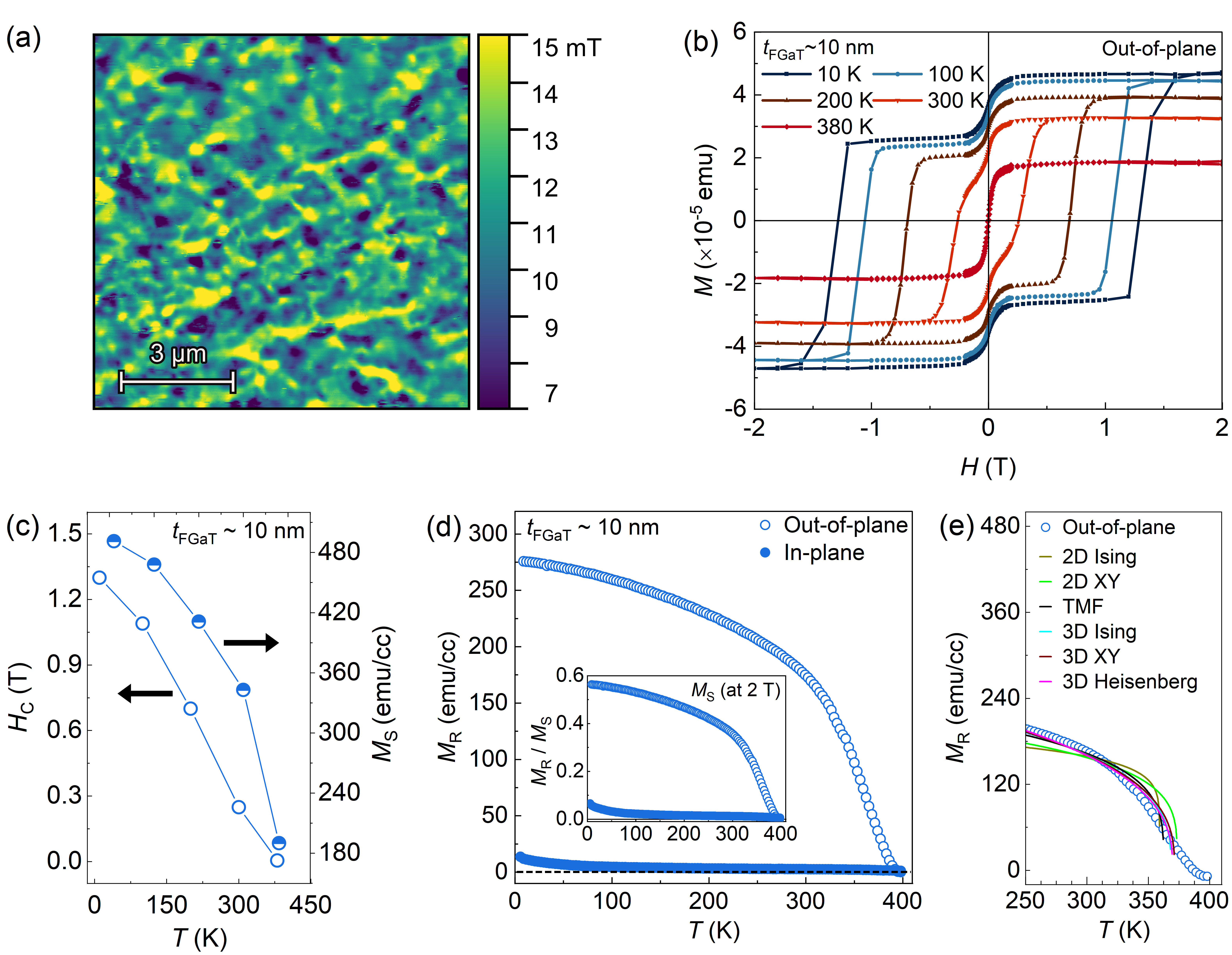}
			\caption { (a) NV field map measured at room temperature for a 32 nm thick FGaT film on graphene/SiC(00.1) using Scanning–NV center microscopy. (b) Magnetization ($M$) versus magnetic field ($H$) hysteresis loop measured at different temperatures for a 10~nm thick FGaT film on graphene/SiC(00.1). (c) Extracted values of $H_{\rm C}$ and $M_{\rm S}$ as a function of temperature. (d) Remanent magnetization, $M_{\rm R}$ as a function of temperature for both in-plane ($H \parallel ab$) and out-of-plane ($H \parallel c$) configurations, the insets represents the corresponding $M_{\rm R}/M_{\rm S}$. (e) $M_{\rm R}$ as a function of temperature for the out-of-plane ($H \parallel c$) configuration plotted together with fittings obtained with Eq.~\eqref{MT-fit} using different critical exponents.}
						\label{fig4}	
		\end{center}
\end{figure}
\subsection{Magnetic properties}
Single Scanning Nitrogen vacancy (NV) center-based microscopy was used to probe the magnetic domain structure of the MBE-grown FGaT films. Fig.~\ref{fig4}(a) shows an NV field map collected for the 32 nm thick FGaT film. This magnetic field map was measured at 300~K, indicating the formation of a stable magnetic domain structure at room temperature. Static magnetization measurements were performed using a SQUID (superconducting quantum interference device) magnetometer to investigate the magnetic properties of FGaT films, namely coercive field ($H_{\rm C}$), Curie temperature ($T_{\rm C}$), and saturation magnetization ($M_{\rm S}$). In Fig.~\ref{fig4}(b), the magnetization hysteresis loop for a 10 nm-thick FGaT film was measured at different temperatures by sweeping the magnetic field in out-of-plane ($H \parallel c$) ranging between +2 and -2 T. A strong linear diamagnetic contribution signal originating from the graphene/SiC(00.1) template was subtracted from the curves. Besides a large coercivity up to room temperature, a small magnetic contribution also adds and leads to a two steps switching feature in the hysteresis loop. This additional loop feature with negligible $H_{\rm C}$ could be related to the existence of multiple magnetic domains within the ferromagnetic layer~\cite{liu_wafer-scale_2017, wu_robust_2024} or due to partial surface oxidation of FGaT.~\cite{kim2019antiferromagnetic}
Further investigations are needed to clarify this aspect. In Fig.~\ref{fig4}(c), the extracted values of $H_{\rm C}$ and $M_{\rm S}$ are plotted as a function of temperature. The large value of $H_{\rm C}$ and $M_{\rm S}$, in particular at 10~K, indicates a strong PMA for the MBE-grown FGaT, which is also confirmed by hysteresis loops obtained via magneto-transport and XMCD measurements (see discussions later). 

SQUID also allowed us to study the temperature dependence of the remanence magnetization ($M_{\rm R}$). For this measurements, we first applied a large positive field of +5 T to magnetize the FGaT film completely. After the field was set to 0~T in linear mode, the $M_{\rm R}$  was measured by warming up the temperature from 5 K to 400 K. Fig.~\ref{fig4}(d) shows the results obtained by warming-up the temperature from 5 to 400 K in both in-plane ($H \parallel ab$) and out-of-plane ($H \parallel c$) field configurations. For the out-of-plane case, it is observed that the $M_{\rm R}$ value reduces only $\approx35\%$ (from 276 emu/cc to 180 emu/cc) from 5~K to 300~K, which indicates a good thermal stability of the out-of-plane magnetization. The inset plot shows the normalized $M_{\rm R}/M_{\rm S}$, the $M_{\rm S}$ value used here is taken for 2~T field. Considering the extrapolation of the out-of-plane ($M_{\rm R}$) to zero, a high $T_{\rm C}$ around 390 K can be deduced. In addition, the $T_{\rm C}$ value was also estimated by fitting the out-of-plane $M_{\rm R}-T$ curve shown in Fig.~\ref{fig4}(d) using the Eq.~\eqref{MT-fit} below:~\cite{ribeiro_large-scale_2022}  

\begin{equation}
    M_{\rm R}=M_{\rm 0}\Big(1-\frac{T}{T_{\rm C}}\Big)^{\beta},
    \label{MT-fit} 
\end{equation}

where $\beta$ is the magnetic critical exponent for which we considered different models, namely the 3D Heisenberg, 3D XY, 3D Ising, tricritical mean field (TMF), 2D XY, and 2D Ising. The extracted values of $T_{\rm C}$ and $\beta$ are found to be $371\pm0.7$ K and 0.365, respectively. The curve shows the best fit when $\beta$ from the 3D Heisenberg model is employed, indicating that the 10 nm-thick FGaT film already acts as a 3D magnetic system despite a relatively small thickness of only 6 unit cells of FGaT [see Fig.~\ref{fig0}(d)]. A small gradual decline
or “tail-like” behavior of the $M_{\rm R}$ value near the transition temperature was also observed similar to  other
2D ferromagnets.~\cite{gong_discovery_2017, may2016magnetic, zhang2021room} In any case, the $T_{\rm C}$ obtained from the SQUID investigations confirm that ferromagnetic order persists well above 350 K in our MBE-grown films, which is comparable or higher to what has been shown for bulk single crystals (and flakes),~\cite{wang_hard_2024, zhang_above-room-temperature_2022} as well as MBE films grown on mica.~\cite{wu_robust_2024}

\subsection{Anomalous Hall Effect (AHE)}
\begin{figure}[t!]
	\begin{center}
   \includegraphics[width=0.92\columnwidth]{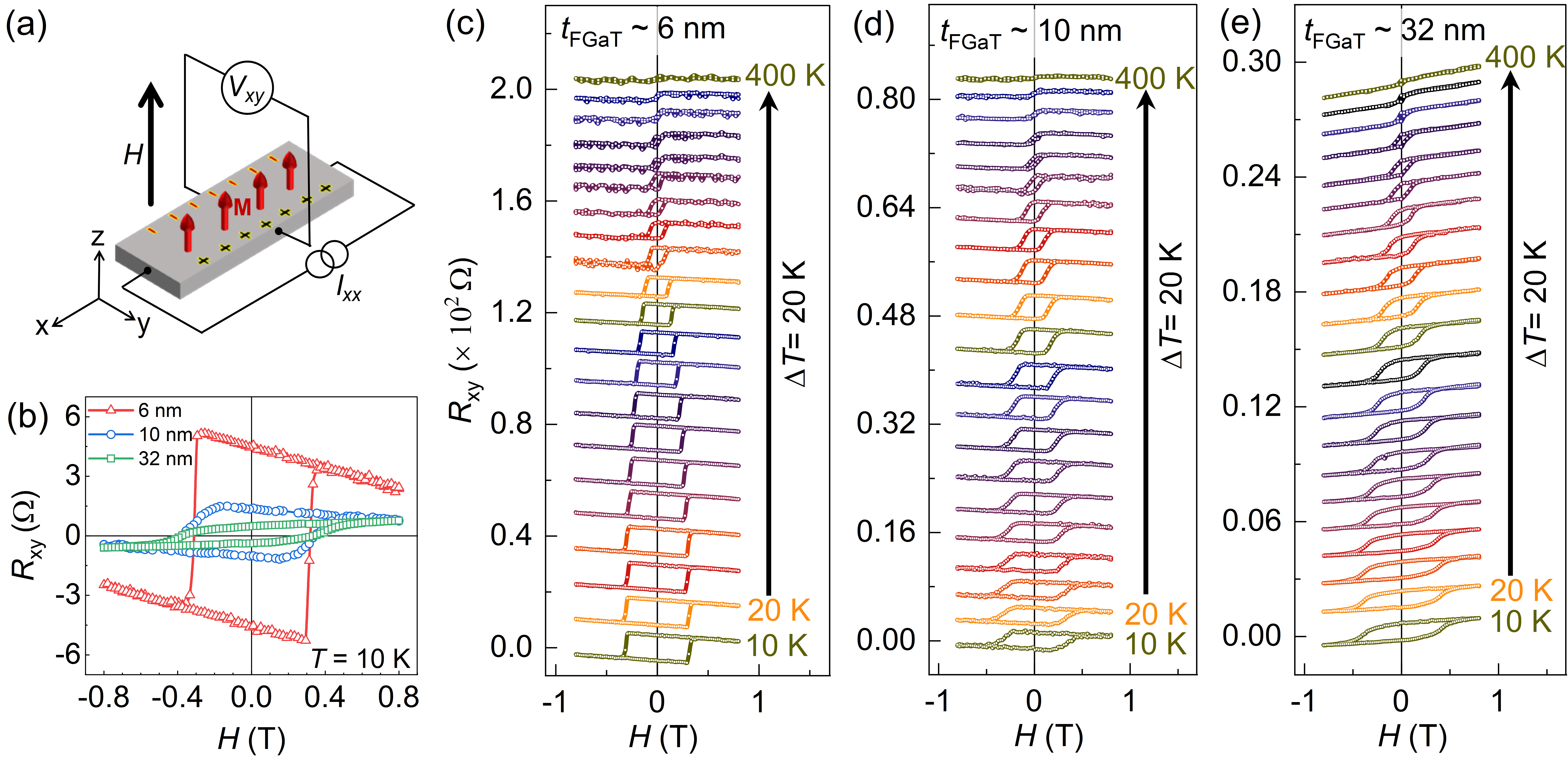}
		\caption {(a) Schematic of the magneto-transport measurement where the Hall voltage ($V_{\rm xy}$) is measured along the \textit{y-}direction while the direct current ($I_{\rm xx}$) flows in the \textit{x-}direction and the external magnetic field ($H$) is swept along the \textit{z-}direction. (b) Transverse resistance ($R_{\rm xy}$) versus external magnetic field ($H$) hysteresis loops for 6~nm,~10~nm~and~32~nm thick FGaT films on graphene/SiC(00.1) measured at 10 K. The strength of coercive field ($H_{\rm C}$) and saturation $R_{\rm AHE}$ is indicated in the same plot. The temperature dependent $R_{\rm xy}$ was measured from 10 K to 400 K at steps of 20 K for (c) 6~nm, (d) 10~nm~and (e) 32~nm thick FGaT films on graphene/SiC(00.1).}
		\label{fig5}	
	\end{center}
\end{figure}

Temperature-dependent magneto-transport measurements were performed for FGaT/graphene heterostructures with different FGaT thicknesses ($t_{\rm FGaT}$= 6 nm, 10 nm, and 32 nm) as shown in Fig.~\ref{fig5}(a).  In ferromagnetic materials, the transverse Hall resistance ($R_{\rm xy}$), under the application of a perpendicular magnetic field ($H$), is given by Eq.~\eqref{AHE-fit}:~\cite{hurd_hall_1972, khan2022intrinsic}
\begin{equation}
R_{xy} = R_{\rm OHE} + R_{\rm AHE},
 \label{AHE-fit} 
\end{equation}
where the first and second terms represent the contribution from the ordinary Hall effect (OHE, $R_{\rm OHE}=R_{0}H$) and the anomalous Hall effect (AHE, $R_{\rm AHE}=\mu_0 R_{\rm S}M_{\rm Z}$), respectively. $R_0$ and $R_{\rm S}$ are the coefficients of ordinary and anomalous Hall resistance, respectively;  $\mu_0$ is the vacuum permeability and $M_{\rm Z}$ is the out-of-plane magnetization. Fig.~\ref{fig5}(b) displays $R_{\rm xy}$  measured at 10 K for samples having  $t_{\rm FGaT}$= 6 nm, 10 nm, and 32 nm. For all cases, the AHE is detected as a square-shaped hysteresis loop. In contrast to double loop $M-H$ feature obtained from SQUID , we observed a single loop shape AHE hysteresis for all the thickness, which confirms the stoichiometric FGaT with a robust PMA~\cite{wu_robust_2024}. Furthermore, the AHE is superimposed on an OHE contribution that has a linear field dependence due to the metallic nature of both the FGaT and the underlying graphene~\cite{lopes_above_2024, lv_large-area_2023} (see the discussions later). Furthermore, the coercive fields ($H_{\rm C}$) are very similar for all samples, while the larger value of $R_{\rm AHE}$ is measured for the thinnest FGaT film. This expected behavior of exhibiting more pronounced $R_{\rm AHE}$ with lower thickness has previously been observed, not only for FGaT,~\cite{wu_robust_2024, wang_hard_2024} but also FGT.~\cite{roemer_robust_2020}

The temperature dependence of $R_{\rm xy}(H)$ is shown in Fig~\ref{fig5}(c)-(e) for the FGaT samples with three different thicknesses. One can see that the AHE vanishes only at 400 K. This confirms the existence of ferromagnetic order with PMA at and above room temperature, consistent with the NV-center and SQUID results. The temperature dependence of $R_{\rm AHE}$ and $H_{\rm C}$ is illustrated in Fig.~6(a) and (b), respectively. Considering the extrapolation of both $R_{\rm AHE}$ and $H_{\rm C}$ quantities to zero, we tentatively conclude that the transition temperature $T_{\rm C}$ of the FGaT films is around 400 K. The value of $T_{\rm C}$ obtained for our films is larger than what has recently been reported for FGaT MBE films grown on GaAs~\cite{wang_centimeter-scale_2024}, but similar to FGaT films (grown on mica)~\cite{wu_robust_2024} and exfoliated flakes~\cite{zhang_above-room-temperature_2022} of comparable thicknesses. 

\begin{table}[b!]
\centering
\caption{\label{FeSn_MH-table}
The value of $R_{\rm AHE}$, $H_{\rm C}$, and $n_{\rm e/h}$ measured at temperature 10~K and 300~K for three different thicknesses of FGaT and graphene/SiC template$^a$;}
\renewcommand*{\arraystretch}{1.6}
\begin{tabular}{| c | c  | c |  c | c |  c | c |} 
\hline
Sample thickness & \multicolumn{2}{c |}{$R_{\rm AHE} $($\Omega$)} & \multicolumn{2}{c |}{$H_{\rm C}$~(T) }& \multicolumn{2}{c |}{$n_{\rm e/h}$~($\times10^{14}/{\rm cm}^{2}$) }\\
\cline{2-7}
 & $10$~K & $300$~K & $10$~K & $300$~K & $10$~K & $300$~K \\
\hline 
FGaT (6 nm) & 4.75 & 2.40 & 0.31  & 0.05 & -2.03 & -2.32\\ 
FGaT (10 nm)& 1.24 & 1.31 & 0.34  & 0.06 & -6.29& -6.51\\ 
FGaT (32 nm)& 0.45 & 0.62 & 0.41  & 0.10 & +29.4& +8.71\\ 
\hline
\end{tabular}
\label{tabp}\\
\footnotesize{$^a$ The value of $n_{\rm e/h}$ for graphene/SiC template is found to be $-0.19\times10^{14}/{\rm cm}^{2}$.}
\end{table}
\begin{figure}[t!]
	\begin{center}
		\includegraphics[width=0.92\columnwidth]{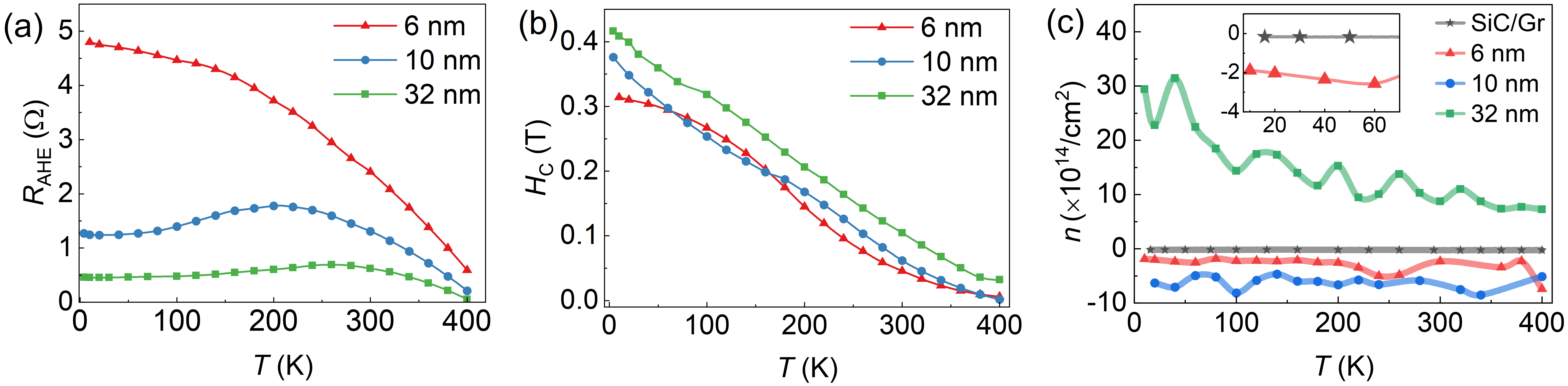}
		\caption {(a) Anomalous Hall resistance $R_{\rm AHE}$ as a function of temperature for FGaT films with different thicknesses.  (b) The dependence of the extracted value of coercive field $H_{\rm C}$  from the transverse resistance contribution as a function of temperature. The solid lines are guides to the eye. (c) Dependence of the hole/electron carrier density on the temperature for FGaT films with different thicknesses and also the pristine graphene/SiC(00.1) template denoted by a star symbol. The average value of electron/hole carrier density ($n_{\rm e/h}$) was determined from the plot, the inset in (c) represents the zoom value for graphene/SiC template compared to 6 nm thick FGaT film.}
		\label{fig6}	
	\end{center}
\end{figure}
Finally, using the relation $R_0=\frac{1}{n_{e/h}|q|}$, the charge carrier type [$q$: electron/hole $(e/h)$] and carrier density ($n_{e/h}$) were determined for the three different thicknesses. Fig.~\ref{fig6}(c) shows the extracted value of $n_{e/h}$ as a function of temperature from 400~K down to 10~K. For the 32 nm thick FGaT film, we observed an increase in the \textit{p}-type carrier density with the decrease in temperature. It is believed that the electronic band structure of FGaT is similar to FGT, only by considering the FGaT structure doped with 1 hole per
formula unit.~\cite{park_controlling_2020, lee_electronic_2023} Therefore, our 32 nm thick FGaT film shows hole-type conductive behavior over the whole temperature range of 10~K-400~K, which is consistent with previous reports.~\cite{zhang_above-room-temperature_2022} In contrast, for the heterostructures containing 6 nm and 10 nm FGaT films, an \textit{n}-type transport behavior was measured over the whole temperature range. We attribute this change in the carrier type in our thinner films to a larger contribution of the underlying graphene layer to the electronic transport. Graphene on SiC(00.1) is known to be intrinsically \textit{n}-type doped.~\cite{schumann_anisotropic_2012, ristein_origin_2012} and to act as a high-mobility transport channel when combined into vdW heterostructures with 2D ferromagnetic metals such as Fe$_3$GeTe$_2$ and Fe$_{5-x}$GeTe$_2$ \cite{lopes_large-area_2021, lv_large-area_2023} Thus, while the transport behavior in thickest FGaT/graphene sample is dominated by FGaT, an enhanced contribution from graphene with two-transport channel occurs in the samples with reduced FGaT thickness. In this case, the value and sign of the slope associated with the OHE will depend on the relative conductivity of both FGaT and graphene and will not correspond to the value of a single material. The extracted values of $n_{\rm e/h}$ measured at 10 K and 300 K for three different thicknesses of FGaT/graphene heterostructure are summarized in Table~\ref{tabp} together with the values for  $R_{\rm AHE}$ and $H_{\rm C}$. Note that the value of $n_{\rm e/h}$ measured for a bare graphene/SiC(00.1) template is mentioned as a footnote of Table~\ref{tabp} for reference. 

\subsection{X-ray absorption spectra (XAS) and X-ray magnetic circular dichroism (XMCD).}
The magnetic properties of FGaT thin films were also studied using XMCD. The measurements were performed on 6 and 10 nm thick FGaT films at normal incidence (NI, $\theta = 90^\circ$) at the Fe $L_{2,3}$ absorption edges. 
\begin{figure}[b!]
	\begin{center}
		\includegraphics[width=0.92\columnwidth]{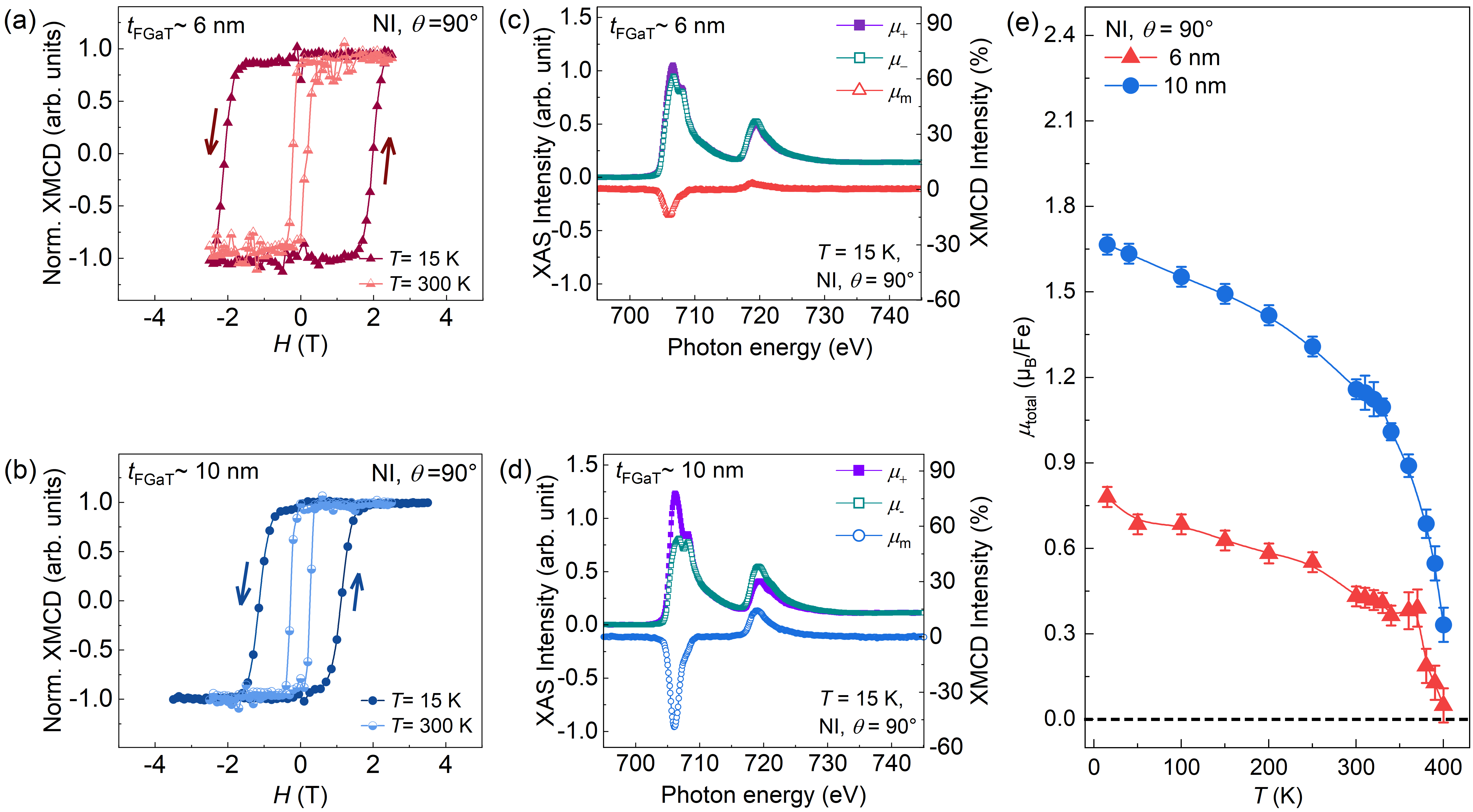}
		\caption{XMCD hysteresis loops probed at Fe $L_{3}$ edge were measured at temperatures of 15~K and 300~K in NI-mode ($\theta=90^\circ$, i.e. the external magnetic field is swept along the \textit{c}-axis of the FGaT layer) for (a) 6 nm-thick FGaT film and (b) 10 nm-thick FGaT film. The XAS and XMCD plots measured at 15~K in NI-mode for (c) 6 nm-thick FGaT film and (d) 10 nm-thick FGaT film. (e) The closed triangle and circle symbols represents the total magnetic moment ($\mu_{\rm total}$) plotted as a function of temperature for 6 nm and 10 nm thick FGaT, respectively, measured in NI-geometry.
        }
		\label{fig7}	
	\end{center}
\end{figure}
Figure~\ref{fig7}(a) and (b) show the XMCD hysteresis loops measured in NI mode at two different temperatures, 15~K and 300~K. These XMCD loops were taken at a photon energy of 706.7~eV (corresponds to the $L_3$ absorption edge of Fe$^{+2}$), along with a pre-edge taken at 698.0~eV for background. At 15~K (300~K), the values of $H_{\rm C}$ for the 6 nm and 10 nm FGaT films were found to be $\sim2$~T ($0.2$~T) and $\sim1.15$~T ($0.25$~T), respectively. These values are in good agreement with the results obtained via SQUID. The large $H_{\rm C}$ values for both films, in particular at 15~K, confirm the strong PMA in our MBE-grown FGaT layers. 
In Fig.~\ref{fig7}(c) and (d), the XMCD spectra ($\mu_{\rm m}$) were extracted by taking the differences of the XAS spectra ($\mu_{+}$ and  $\mu_{-}$) measured at 15 K with a constant magnetic field ($\pm H= 2.5$~T) for both thicknesses. The XAS and XMCD intensities are denoted as $\mu_{+}$, $\mu_{-}$ and $\mu_{m}$ on the left and right axes of the plot, respectively. Similarly, for both samples, the XMCD spectra were obtained over the temperature range from 15~K to 400~K. 
By applying the sum rule anaylysis,~\cite{thole_x-ray_1992, carra_x-ray_1993, chen_experimental_1995} the values of effective spin magnetic moment ($\mu_{S}^{\rm eff}$), orbital magnetic moment ($\mu_{l}$), total magnetic moment ($\mu_{\rm total}$) were determined (see supplementary file S2). In Fig.\ref{fig7}(e), the extracted values of $\mu_{\rm total}$ for both film thicknesses are plotted as a function of temperature. At 15~K, the values of $\mu_{S}^{\rm eff}$, $\mu_{l}$, and $\mu_{\rm total}$ for the 10~nm [6~nm] thick FGaT film were found to be 1.53~$\mu_{\rm B}$/Fe [0.74~$\mu_{\rm B}$/Fe], 0.14~$\mu_{\rm B}$/Fe [0.04~$\mu_{\rm B}$/Fe], and 1.67~$\mu_{\rm B}$/Fe [0.78~$\mu_{\rm B}$/Fe], respectively. At 300~K, these values decreased to 1.13~$\mu_{\rm B}$/Fe [0.45~$\mu_{\rm B}$/Fe], 0.03~$\mu_{\rm B}$/Fe [0.02~$\mu_{\rm B}$/Fe], and 1.16~$\mu_{\rm B}$/Fe [0.47~$\mu_{\rm B}$/Fe], respectively. Notably, the value of $\mu_{\rm total}$ at 300~K for the 10~nm film (1.16~$\mu_{\rm B}$/Fe) is in excellent agreement with values previously reported for FGaT bulk crystals using SQUID magnetometry.~\cite{zhang_above-room-temperature_2022} The ratio of orbital to spin magnetic moment ($\mu_l/\mu_S^{\rm eff}$) was determined to be 0.092 [0.058] at 15~K and 0.027 [0.049] at 300~K for the 10~nm [6~nm] thick FGaT films. The non-zero $\mu_{\rm total}$ for both thicknesses up to 400~K as shown in Fig.~\ref{fig7}(e), reveals a stable ferromagnetic ordering of Fe atoms in the FGaT layer sustaining a robust PMA above room temperature. This temperature-dependent behavior of $\mu_{\rm total}$ for both FGaT thicknesses further confirms the high $T_{\rm C}\sim 400$~K consistent with the results obtained from $M-T$ curves by SQUID and $R_{\rm AHE}-T$ curves obtained via magneto-transport. Therefore, these results on high-quality vdW-based heterostructures are pivotal for advancing next-generation non-volatile memory technologies, including magnetic random-access memory (MRAM) and skyrmion-based data storage.~\cite{ahn_2d_2020, liu_magnetic_2024} 

\section{Conclusion}
We have demonstrated the high-quality, large-scale epitaxial growth of 2D ferromagnetic Fe$_{3}$GaTe$_{2}$ films directly on graphene/SiC(00.1) substrates. Structural characterization confirms the excellent crystalline quality, homogeneous surface, and sharp interface of the FGaT/graphene van der Waals heterostructure. Magnetic and transport measurements reveal robust perpendicular magnetic anisotropy (PMA) with a high Curie temperature reaching up to 400~K. These findings are further substantiated by XMCD measurements, where perpendicular magnetic moments were probed under normal incidence geometry. Additionally, the spin, orbital, and total magnetic moment contributions were extracted from X-ray absorption spectroscopy performed over a wide temperature range (10 K to 400 K).  At 300~K, we obtained the $\mu_{\rm total}$ to be $1.16~\mu_{\rm B}$/Fe  for our 10~nm thick FGaT film, which is consistent with the previous reports on FGaT bulk crystals. Our findings underscore the potential of FGaT large-scale epitaxial films as a platform for high-performance spintronic devices based on 2D heterostructures, with potential applications in future data storage, processing technologies, and quantum computing.

\section{Experimental Method}
\subsection{Sample preparation}
We grew epitaxial graphene on semi-insulating 4H-SiC(00.1) substrates using the SiC surface graphitization method.~\cite{heilmann_influence_2020} Before this, the substrates were chemically cleaned under ultrasonication using n-butyl acetate, acetone, iso-propanol and later dried with N$_2$-gas. This was followed by an annealing in a forming gas environment consisting of 95\% Ar and 5\% H$_2$ gases at 1400 $^\circ$C for 15 min, to obtain smooth, well-defined stepped surfaces. Then, the synthesis of smooth layers of epitaxial graphene was realized in an atmosphere of Ar gas at a temperature of 1600 $^\circ$C for 15 min, similar to our previous work.~\cite{lopes_large-area_2021, lv_large-area_2023} 

The high-quality large-scale epitaxial Fe$_{3}$GaTe$_{2}$ thin films were successfully grown on graphene/SiC(00.1) templates, in ultra-high vacuum (UHV) conditions using MBE. Before the film growth, the templates were degassed at 450~$^\circ$C for 1 hour in the UHV chamber to remove surface contaminants. High-purity elemental Fe, Ga, and Te were evaporated from separate effusion cells. Their temperatures were chosen aiming at realizing films with the desired composition of Fe$_{3}$GaTe$_{2}$, and the fluxes for each element were obtained by measuring the beam equivalent pressure employing a pressure gauge. The growth chamber was maintained at a base pressure below 1$\times$10$^{-10}$ mbar and a working pressure below 1$\times$10$^{-9}$ mbar. An optimized substrate temperature of 370~$^\circ$C was chosen to get a high-quality crystalline phase of FGaT film. After reaching the desired thickness, ranging from 6 to 32~nm, the samples were allowed to cool down to room temperature (RT).

Later, at RT, a 5-7~nm Te capping layer was deposited in order to minimize the oxidation of the FGaT films upon air exposure. For some of the samples, an additional Pt capping layer was deposited in-situ, also at RT, on top of the Te capping layer to avoid oxidation of the Te layer. The entire growth process was monitored using in-situ RHEED to assess the crystalline quality of the growing films. A low growth rate of 0.17~nm/min was employed during the FGaT growth. 

\subsection{Composition and Structural characterization}
The accurate value of the chemical composition of FGaT films was determined with the help of the Rutherford backscattering (RBS)  technique, using 1.7~MeV He$^{+}$ ions with a scattering angle of 170$^\circ$. From the RBS measurements, the chemical composition of Fe:Ga:Te was found to be around 3.15:1:1.47, revealing thus a small Fe excess and Te deficiency in comparison to stoichiometric FGaT. The surface morphology and the film thickness were examined by atomic force microscopy (AFM), conducted in the standard tapping mode. For the STEM measurements, the lamellae were prepared using the Thermo Fisher Helios 5 UX focused Ion beam (FIB) system, which is equipped with a Ga$^{+}$ ion beam source operating in the energy range from 1 to 30 keV. Using a nano-manipulator, the cross-sectional lamella was lifted out according to standard procedures and transferred to a grid. The lamella was then thinned to the desired shape and thickness, and in a final step, polished with a low-energy ion beam to minimize the surface damage caused during FIB preparation. The lamellae were then immediately inserted into the TEM for structural observation to minimize oxidation of the iron-containing layer. The STEM measurements were carried out in a Jeol 2100 F field emission electron microscope at an acceleration voltage of 200 kV. The individual thickness of each layer in the film stack was further characterized by the X-ray reflectivity (XRR). Both XRR and out-of-plane XRD ($\theta-2\theta$ scan) were performed using a PANalytical X'Pert Pro MRD diffractometer with Cu-$K_\alpha$ radiation ($\lambda$ = 1.5418 \AA). To investigate the in-plane crystallographic properties, grazing incidence diffraction (GIXRD) was performed at the synchrotron beamline BM25B of the European Synchrotron Radiation Facility (ESRF) in Grenoble. A monochromatized X-ray beam was used with a photon energy of 18\,keV. A grazing incidence angle close to the one of total external reflection makes this technique extremely surface sensitive.

\subsection{Magnetization measurement}
The static magnetization measurements were performed using a Quantum Design MPMS3 SQUID magnetometer at different temperatures between 10~K and 400~K by sweeping the applied magnetic field up to 2~T. Magnetization vs. temperature measurements were performed in both in-plane and out-of-plane configurations for the same sample pieces. Magnetic field sweeps were performed in a persistent mode, the temperature sweeps were performed at a rate of 5~K per minute. 

The stray field image of the magnetic domain was acquired on a commercially available single Nitrogen-Vacancy (NV) scanning microscope (QZabre - QST).

\subsection{Anomalous Hall measurement}
Anomalous Hall measurements were conducted on a rectangular strip [0.5~cm $\times$ 1~cm] piece of the Fe$_{3}$GaTe$_{2}$ on graphene/SiC(000), which was bonded to a chip carrier using aluminum contact wires. A constant current of 100 mA was applied along the longest side of the piece in the x-direction. The transverse voltage ($V_{\rm xy}$) was recorded perpendicular to the direct current ($I_{\rm xx}$) flow using two centrally positioned contacts on the sample strip along the y-axis. Experiments were conducted across a wide temperature range, spanning from 10~K to 400~K, in a high-vacuum environment with pressures maintained between 10$^{-6}$~mbar. The external magnetic field was swept between $\pm0.8$~T perpendicularly to the sample surface (along the \textit{c}-axis).

\subsection{XAS and XMCD measurements} 
The XAS measurements were conducted at the VEKMAG end-station on the PM2 beamline at BESSY II. This facility provides a vector magnetic field of up to 9~T along the beam direction, 2 Tesla in the horizontal plane, and 1 Tesla in all directions, within a temperature range of 2 K to 500 K. The spectra were recorded using the total electron yield (TEY) method, which measures the drain current as a function of x-ray photon energy. TEY, being surface-sensitive with an electron escape depth of $\sim$3 nm, selectively provides information on surface magnetic properties.~\cite{luo_x-ray_2019} A magnetic field of up to $\pm$ 2.5 T was applied along the beam direction to magnetically polarize the sample during the XMCD experiments. The XMCD signal was obtained by calculating the difference between the XAS signals measured under two opposite magnetic fields.~\cite{silinskas_self-intercalation_2024} The sum rule analysis were performed on integral signal of XAS and XMCD spectra's to calculate the contribution from spin and orbital magnetic moment (see supplementary S2). 

\section{Acknowledgement}
The authors would like to thank H.-P. Schönherr, C. Herrmann, and C. Stemmler for their dedicated maintenance of the MBE system and C. Matzeck for FIB lamella preparation for TEM measurement. Furthermore, the authors appreciate the critical reading of the manuscript by Van Dong Pham. They also acknowledge the provision of beamtime under the project HC-5796 at the European Synchrotron Radiation Facility (ESRF), located in Grenoble (France) and PHARAO station at BESSY II, Helmholtz-Zentrum Berlin. J.M.J.L. acknowledges financial support from the FLAG-ERA (MagicTune project) and the German Research Foundation, DFG (project 533948427). XAS/XMCD measurements were carried out at the VEKMAG station at the BESSY II electron storage ring operated by the Helmholtz-Zentrum Berlin für Materialien und Energie. The RBS measurements were carried out at IBC  at the Helmholtz-Zentrrum Dresden-Rossendorf e.V., a member of the Helmholtz Association.

\section{\large Supporting Information for: Above room temperature ferromagnetism in large-area epitaxial Fe$_3$GaTe$_2$/graphene van der Waals heterostructures}
\subsection{S1: Composition analysis: Rutherford back scattering (RBS)}
\begin{figure*} [htbp]
\centering
\includegraphics*[width=0.9\columnwidth]{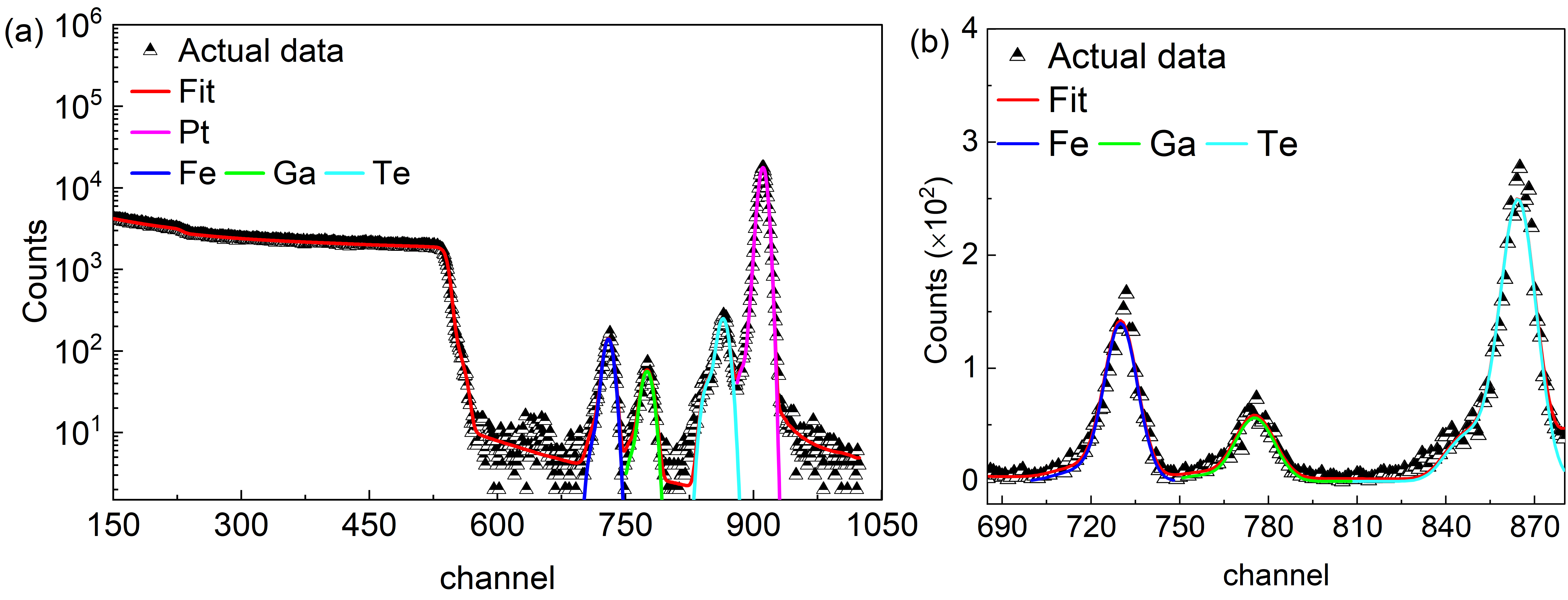}
\caption{\label{S1: RBS } (a) Full RBS spectra for all the elements contained in Pt/FGaT/Gr/SiC heterostructures and (b) the zoomed image corresponds to the spectra for the elements of Fe, Ga, and Te. The symbols and lines represent the experimental values and fit to the spectra.}
\end{figure*}

\subsection{S2: XAS and XMCD sum rule analysis.}
\begin{figure*}[t!]
\centering
\includegraphics*[width=0.7\columnwidth]{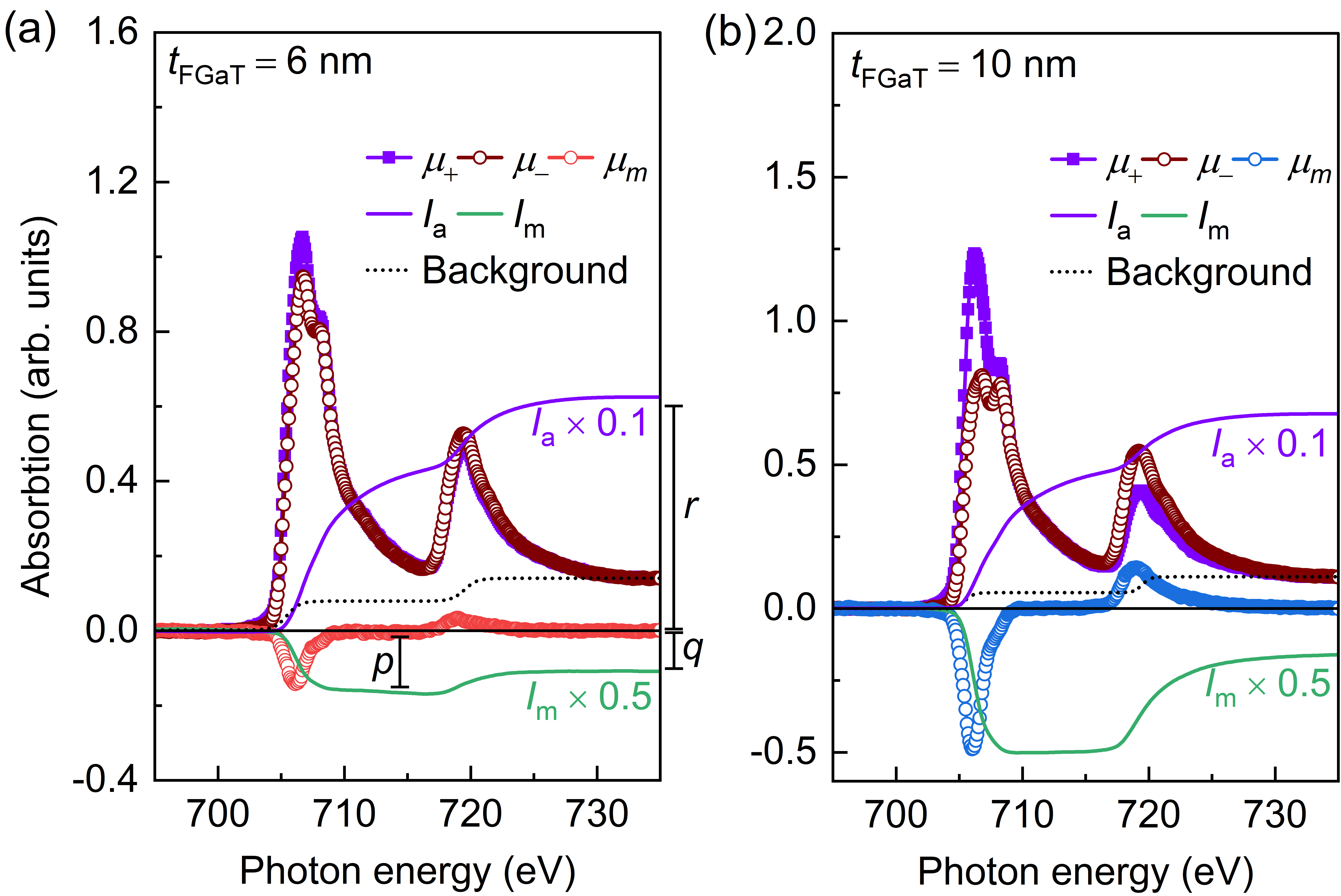}
\caption{\label{S2:sumrule} XAS, XMCD, integrals and the background spectra's of Fe-edge, measured at 15~K, with the sum rule analysis for (a) 6 nm and (b) 10 nm thick FGaT/graphene heterostructure. }
\end{figure*}

To extract the contribution of spin ($\mu_{s}$) and orbital ($\mu_{l}$) magnetic moments in the FGaT thin film, we performed sum-rule analysis on the XAS and XMCD spectra at the Fe-$L_{2,3}$ edges. In Fig.~\ref{S2:sumrule}(a)\&(b), the value of $\mu_{l}$ was obtained from the XMCD integral ($I_{m}$) over the $L_{3}$ edge of Fe, while the effective $\mu_{s}^{\rm eff}$ was calculated from a combination of XMCD integral ($I_{ m}$) and XAS integral ($I_{a}$) over both the $L_{2}$ and $L_{3}$ edges. To accurately determine the values of $\mu_{s}^{\rm eff}$ and $\mu_{l}$, we use the different absorption and dichroic integrals formulae below~\cite{chen_experimental_1995, backes2024valence, lv_large-area_2023}:
\begin{equation}
\mu_{s}^{\rm eff}=-2\langle S_{\rm z}\rangle-7\langle T_{\rm z}\rangle
= -\frac{N_{\rm h}}{P{\rm cos}(\theta)}\frac{(6p-4q)}{r}
\end{equation}
and
\begin{equation}
\mu_{l}=-\langle L_{\rm z}\rangle = -\frac{N_{\rm h}}{P{\rm cos}(\theta)}\frac{4q}{3r}
\end{equation}
where, 
\begin{equation*}
p=\int_{L_3}(\mu_m),~q=\int_{{L_2+L_3}}(\mu_m),~{\rm and}~r=\int_{{L_2+L_3}} \left [\frac{(\mu_+ + \mu_-)}{2}-background\right]
\end{equation*}
Here, $\mu_m(=\mu_- - \mu_+)$ is the XMCD signal. The symbol $\mu_-$ and $\mu_+$ refer to XAS spectra taken under $\pm H$ magnetic field. $I_a$ and  $I_m$ are the integrated plots of XAS and XMCD plots. For both cases, the background signal is known to be originated from the double step function. This background signal was already substrate for the integrals $I_a$ of XAS plot as shown in Fig.~\ref{S2:sumrule}(a)\&(b). $L_z$, $S_z$, and $T_z$ are the atomic orbital, spin magnetic moment, and magnetic dipole operator, respectively. $N_h$ is the number of holes in $3d$ shell (for Fe in valence state of +2, the value 3.4 is used in sum-rule), $P$ is the degree of circular polarization of X-ray beam (77\% polarization of photon energy) and $\theta$ is the angle between the X-ray beam direction and the magnetization direction.

\providecommand{\noopsort}[1]{}\providecommand{\singleletter}[1]{#1}%

\end{document}